\documentclass[english,american]{epl}
\usepackage{times}
\usepackage[T1]{fontenc}
\usepackage[latin1]{inputenc}
\usepackage{graphicx}
\usepackage{amssymb}

\makeatletter

\newcommand{\noun}[1]{\textsc{#1}}
\providecommand{\boldsymbol}[1]{\mbox{\boldmath $#1$}}

\usepackage{float}
\usepackage{amsmath}
\usepackage{amssymb}

\DeclareMathOperator{\Pre}{In}

\usepackage{babel}
\makeatother

\title{Does dynamics reflect topology in directed networks?}

\author{Marc Timme}

\institute{Center for Applied Mathematics, Cornell University,\\
Theoretical and Applied Mechanics -- Kimball
Hall, Ithaca, NY 14853, USA;\\
 Max Planck Institute for Dynamics and Self-Organization (MPIDS), and 
Bernstein Center for Computational Neuroscience (BCCN) Göttingen, Bunsenstr. 10, 37073 Göttingen, Germany.}

\begin{document}

\maketitle

\begin{abstract}
We present and analyze a topologically induced transition from ordered,
synchronized to disordered dynamics in directed networks of oscillators.
The analysis reveals where in the space of networks this transition
occurs and its underlying mechanisms. If disordered, the dynamics
of the units is precisely determined by the topology of the network
and thus characteristic for it.  We develop a method to predict
the disordered dynamics from topology. The results suggest a new
route towards understanding how the precise dynamics of the units
of a directed network may encode information about its topology. 
\end{abstract}

Networks of interacting units prevail in a variety of systems, ranging
from gene regulatory networks and neural networks to food webs and
the world wide web \cite{Bornholdt,AlbertEtAl}. A fundamental question
is: What kind of dynamics can we expect given a network of prescribed
connection topology \cite{Strogatz}? Even in networks of known dynamical
units, known type of interactions between them and known topological
details, it is hard to infer which kind of typical collective dynamics
the network will display (cf.\ Refs.~\cite{Strogatz,TimmePRL,Klemm,Stewart}).
If parts of the network exhibit residual symmetries, such as permutation
or translation invariance, some general properties of the dynamics
can be deduced \cite{Stewart}. If, however, no symmetries remain,
it is still an open question how topological factors can control network dynamics. See, e.g., \cite{Zanette,Kori,Arenas} for some interesting recent approaches for phase oscillator networks of specific connectivities.

In this Letter, we study directed networks of phase oscillators that exhibit a mechanism to synchronize and reveal general principles about
how topology controls dynamics: Specifically, in networks with an
invariant state of in-phase synchrony we analyze how the topology
can control the units' dynamics in a neighborhood of synchrony. We
find that such networks, depending on their coarse-scale topological
properties, belong to one of two classes exhibiting very different
long term dynamics: Networks of class I show in-phase synchrony in
which each unit displays identical dynamics, independent of the unit's
\emph{topological identity} \cite{TopID}, i.e.\ independent of where
in the network it is located. Networks of class II, instead of synchrony,
show disordered dynamics. Here, together with the initial condition,
the fine-scale topology precisely controls the dynamics of each unit.
The dynamics therefore strongly depends on where the unit is located
in the network -- its topological identity. We develop a method to predict the disordered dynamics from
the network's topology. Due to their topological origin, both the separation of the ensemble of networks into two unique classes and the specific disordered dynamics realized
by a network appear to be general phenomena and not restricted to the system studied here. 

Let us elaborate these findings. Consider a network of $N$ phase
oscillators $i$ that interact via directed connections. The network
topology is arbitrary and determined by the sets $\Pre(i)$ of those
units $j$ that have input connections to $i$, denoted $j\rightarrow i$.
We analyze a simple, paradigmatic model of interacting periodic oscillators,
the Kuramoto model \cite{Kuramoto,StrogatzKuramoto,Acebron} defined
by\begin{equation}
\frac{d}{dt}\phi_{i}(t)=\omega_{i}+\sum_{j\in\Pre(i)}J_{ij}\sin(\phi_{j}-\phi_{i})\label{eq:KuramotoDE}\end{equation}
 where the phase variable $\phi_{i}(t)\in[0,2\pi)$ (with periodic
boundary conditions) determines the state of unit $i$ at time $t$,
$\omega_{i}$ is its frequency, and $J_{ij}\geq0$ is the strength
of coupling from $j$ to $i$ with $J_{ij}=0$ if there is no connection
$j\rightarrow i$. In order to stress the topological effects, we
neglect inhomogeneities in the dynamical parameters: we consider identical
units $\omega_{i}=\omega$ and homogeneous total input coupling strengths
such that $\sum_{j}J_{ij}=J$ (in all illustrating examples we choose
$J_{ij}=J/k_{i}$ if unit $i$ receives $k_{i}$ input connections
from other units $j$). Without loss of generality, we take $J=1$
in the following. We consider initial states in a neighborhood of
the in-phase synchronous solution to reveal those features that apply
for other oscillator networks as well. 
%
\begin{figure}
\begin{center}\includegraphics[%
  width=90mm]{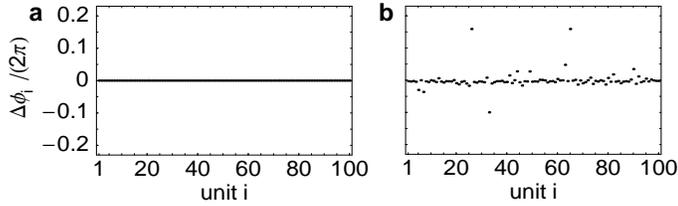}\end{center}

\caption{Synchronization-disorder transition in directed network dynamics.
(a),(b) The long time dynamics of two different random network realizations
of $N=100$ units with the same connection probability $p=0.05$ started
from the same random initial condition. The phase differences $\Delta\phi_{i}:=\phi_{i}-\left\langle \phi_{j}\right\rangle _{j}$
with respect to the average phase $\left\langle \phi_{j}\right\rangle _{j}$
of all units are shown versus the units $i$, plotted relative to
their possible range, $2\pi$.\label{fig:order-disorderN100}}
\end{figure}

Observing the dynamics (\ref{eq:KuramotoDE}) on different topologies,
it was intriguing to find that seemingly similar networks (such as realizations of networks with identical degree distribution) yet displayed very different
dynamics. Consider for instance the long-term dynamics of random networks
in which every connection $j\rightarrow i$ is present with probability
$p$. Several such networks show in-phase synchrony (cf.\ Fig.~\ref{fig:order-disorderN100}a).
Thus, the final states of the units display no information about the
network topology. The units' \emph{topological identity} is hidden. Other networks with identical statistical properties 
display disordered periodic dynamics,
even when initialized in the same state (Fig \ref{fig:order-disorderN100}b).
In such disordered states almost every unit displays a different phase.
It turns out (see below) that the network topology precisely controls
the dynamics of these units. The units' dynamics thus display their
topological identity!

This phenomenon raises a number of questions. In which networks and how 
does the disordered state emerge? 
What determines the individual units' dynamics
if the network displays disorder?

\begin{figure}
\begin{center}\includegraphics[%
  height=115mm,
  angle=-90]{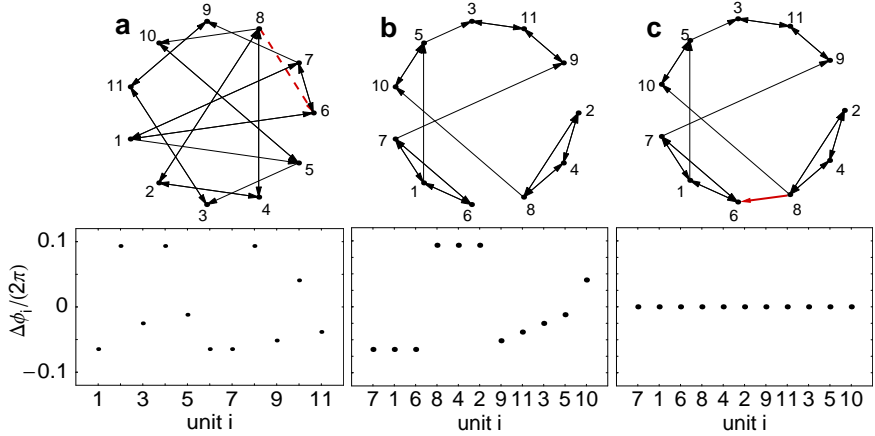}\end{center}

\caption{(color online) Dynamics of small networks ($N=11$) started from
the same random initial condition. The upper column of each panel
displays the directed networks with units labeled $i\in\{1,\ldots,N\}$.
The lower displays the relative phase differences $\Delta\phi_{i}/(2\pi)$
versus $i$. (a) A network with homogeneous in-degree $k_{i}=2$ for
all units $i$ exhibits an irregular asymptotic state. (b) Same network
as in (a) with units with similar phases grouped. (c) A network with
only one more directed edge ($\textsf{{8}}\rightarrow\textsf{{6}})$
{[}red; red dashed edge in (a){]} compared to that of (b) induces
a completely ordered state with identical dynamics of all units. \label{fig:order-disorderN11}}
\end{figure}

To answer these questions, we studied the dynamics of small networks
that exhibit qualitatively the same disordered state (see, e.g. Fig.~\ref{fig:order-disorderN11}a).
We tried to find a systematic dependence of the units' states on the
network topology. As a first step, we ordered the units of the network
(Fig.~2a) such that units with similar phases are displayed at proximate
positions (Fig.~2b). This ordering reveals a division of the network
in terms of its strongly connected components (SCCs) \cite{SCC}.
Whereas there are some units with identical phases, several phases
appear uniquely. In a network with just one more edge (Fig.~2c),
the collective dynamics is completely synchronized such that the topological
identity of all units is hidden (cf.~also Fig.~1a). The above finding
that ordering of the phases seems to reveal information about the
coarse scale network topology led us to hypothesize that the partition
of the network into SCCs is important to understand its dynamics.

To test this hypothesis, we first analyze the dynamics of networks
of arbitrary connectivity in a neighborhood of in-phase synchrony
($\phi_{i}(t)=\phi_{0}(t)$ for all units $i$ and all
times $t$). Disconnected parts of a network can be treated independently,
such that we focus on connected networks here. Sufficiently small
perturbations $\delta_{i}(t):=\phi_{i}(t)-\phi_{0}(t)$ to the synchronous
state satisfy \begin{equation}
\dot{{\delta_{i}}}=\sum_{j\in\Pre(i)}J_{ij}\sin(\delta_{j}-\delta_{i})\label{eq:nonlinearDE}\end{equation}
 for all $i$, which in first order approximation reads $\dot{{\delta_{i}}}=-J\delta_{i}+\sum_{j\in\Pre(i)}J_{ij}\delta_{j}$,
or $\dot{\boldsymbol{\delta}}\doteq M\boldsymbol{\delta}$ in matrix
form, where\begin{equation}
M_{ij}=\left\{ \begin{array}{cl}
-J & \textrm{if }j=i\\
J_{ij} & \textrm{if }j\in\Pre(i)\\
0 & \textrm{if }j\notin\{ i\}\cup\Pre(i)\end{array}\right.\label{eq:Mij}\end{equation}
 are the matrix elements of $M$ and $\boldsymbol{\delta}=(\delta_{1},\ldots,\delta_{N})^{\mathsf{{T}}}$
is the vector of the individual units' perturbations $\delta_{i}$.
This results in the first order period-$T$ map ($T=2\pi/\omega$)
given by \begin{equation}
\boldsymbol{\delta}(T)=A\boldsymbol{\delta}(0)\label{eq:linearmap}\end{equation}
 where the matrix elements of $A=e^{MT}$ satisfy $A_{ij}\geq0$,
reflecting the attractive couplings $J_{ij}\geq0$, and $\sum_{j}A_{ij}=1$
due to time translation invariance of the periodic orbit. 

For networks of arbitrary connectivities, this implies, via the Ger\v{s}gorin disk theorem \cite{Mehta}, that all eigenvalues $\lambda_{i}$ of $A$ satisfy
$|\lambda_{i}|\leq1$. A sufficiently small perturbation to the synchronous state cannot grow (in maximum norm), cf. \cite{TimmePRL}, such that synchrony is 
at least marginally stable. 
Moreover, independent of the network connectivity there
is one eigenvalue $\lambda_{1}=1$ with an eigenvector $\mathbf{{v}_{1}}=(1,1,\ldots,1)$ corresponding to the uniform phase shift. 

If the network is strongly connected \cite{SCC} the Perron-Frobenius theorem \cite{Mehta}
guarantees that the largest
eigenvalue $\lambda_{1}=1$ is unique and all other eigenvalues satisfy $|\lambda_{i}|<1$ for $i\in\{2,\ldots,N\}$. 
This implies that the synchronized
state is asymptotically stable and thus locally attracting. In networks of irregular topology we even often find that the system converges towards it from arbitrary initial conditions.

If the network is not strongly connected it consists of two or more
strongly connected components (SCCs) and the analysis of the asymptotic
dynamics is more involved. For better accessibility of the main points
of this Letter we describe the details of this analysis in the Appendix.
Briefly, for a given network, we first determine the SCCs and the
uni-directional connections among them. Second, we determine the level
structure of this super-network of SCCs (cf.\ Fig.~\ref{cap:topologicalanalysis}). 
Based on this composition analysis we have revealed a number of distinctive
features of the dynamics on directed networks: %
\begin{figure}\begin{center}\includegraphics[%
  width=65mm]{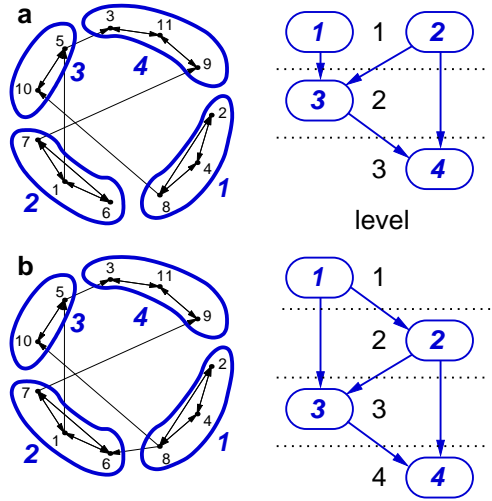}\end{center}

\caption{(color online) Decomposition of networks shown in Fig.\ 2 in terms
of their strongly connected components (SCCs). On the left, the vertices
of the networks are grouped to SCCs $s\in\{1,2,3,4\}$ (large italic
numbers). On the right, the level structure of these components is
shown. (a) Three-level network with two source, cf.\ Fig.\ 2a,b.
(b) One additional link makes it a four-level network with one source,
cf.\ Fig.\ 2c. \label{cap:topologicalanalysis}}
\end{figure}

The ensemble of networks divides into two classes with qualitatively different
long term dynamics (initialized sufficiently close to the in-phase
solution). All networks that have $M=1$ source SCC (which does not receive
any input from other SCCs) belong to class I: this source SCC
is guaranteed to synchronize because it itself is a strongly connected network without further input. Since each unit $i$ performs a local weighted averaging of phases determined by the weights $A_{ij}$ in 
(\ref{eq:linearmap}), all units outside the source component asymptotically converge towards the (common) phase of the units within the (only) source SCC. This result also follows explicitely from the analysis given in the appendix in the 
special case of only one source component in level $\ell=1$ (and no source components in levels $\ell>1$).  It implies that for all networks with one source SCC the local asymptotic dynamics is also in-phase synchrony. 
In contrast, networks having $M \geq 2$ source SCCs (class II) typically show disordered dynamics. These $M$ source SCCs can synchronize independently of each other, creating $M-1$ independent phase differences which result in an $(M-1)$-dimensional continuous family of periodic orbits, that include the synchronous state as only \emph{one} specific orbit. All these orbits are marginally stable, in particular the synchronous state has a basin of attraction of measure zero, such that the dynamics is almost surely disordered. 
For the examples above, we find that the dynamics shown in Fig.~1a originates from a class I network whereas that of Fig.~1b originates from a class II network.  

The composition analysis also reveals how the details of
the topology of the network precisely control its dynamics in the
disordered state: The topological identity of each unit, particularly the fine scale topology of that SCC it is part of, determines the unit's dynamics.
In fact, we can  predict the disordered dynamics on a fine scale: 
Given the initial state $\boldsymbol{\phi}(0)$, we uniquely determine the 
approximate phases of all units recursively
level by level, and hence predict the complete collective dynamics
of the network from the topological identity of their units (see Appendix).
Figure \ref{cap:Prediction} illustrates such a prediction. 
It resembles well the actual dynamics of the units. 

Reversely, partial information about the topology of
the network may be obtained from knowing the disordered dynamics of its 
units: Iterating Eqn. (\ref{eq:phi_no_source}), we obtain 
explicit linear restrictions of the space of all networks from the disordered dynamics by imposing its invariance. So only a lower-dimensional subset of networks is consistent with the phase pattern.

What is the mechanism underlying the transition to topology-induced
disorder? The following description is general; nevertheless
it is instructive to imagine, as an illustrating example, a network
composed of two source SCCs and one sink SCC which receives input
from the other two. A strongly connected network, and thus each source
SCC, synchronizes completely. However,
different of these source SCCs typically converge towards different phases,
that depend on the initial state. If now different units in a downstream SCC are pulled
towards different phases, and there is a complicated pattern of connections
between them within this SCC, the dynamics of all its units
will typically be distinct. In particular, the units' dynamics depend
on the phases of the units in connected upstream SCCs, i.e. indirectly
on the initial state of the network and on the specific topology of
the SCC considered.

\begin{figure}
\begin{center}\includegraphics[%
  width=55mm]{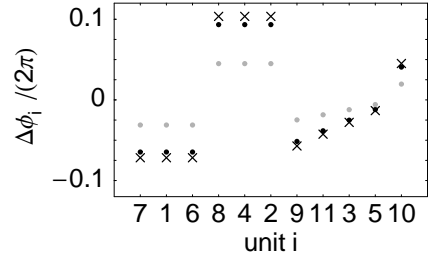}\end{center}

\caption{Prediction of the dynamics in a disordered state (cf. Fig. 2a,b) based
on the composition analysis. The relative phase difference $\Delta\phi_{i}/(2\pi)$
is shown for the grouped units $i$. The linear prediction $(\times)$
of the actual phases ($\bullet$) (based on one intial state) well distinguishes the ordered from
the disordered state (which would be a constant at zero) and even
is a good indicator of the quantitative dynamics of the units.\label{cap:Prediction}
The asymptotic phase dynamics started from a different initial state (gray $\bullet$) illustrates that in this example other initial states yield a pattern that is distinguished from the former pattern only by a real mulitplicative factor (in first order approximation).
}
\end{figure}

All these phenomena appear to be general and are not restricted
to the model system (\ref{eq:KuramotoDE}) considered here. This is due
to the topological origin of the phenomena: First,
the transition line between networks of classes I and II is identical
for various kinds of oscillator networks having an invariant in-phase solution. Second, given an initial state sufficiently close to synchrony, the disorder in the long term dynamics is \emph{characteristic
for the topology} of a network: The linear analysis (see
Appendix) holds as well for all disordered dynamics that are 
topologically equivalent within the class of periodic single-variable 
oscillator networks. We checked
this explicitely for networks of 
(i) Kuramoto oscillators (\ref{eq:KuramotoDE})
with coupling functions different from the sine function and (ii)
spiking neural oscillators where interactions are delayed and mediated
by pulses that occur only at discrete instances of time \cite{TimmePRL,Ernst}.
Although we have no proof of how general these results are beyond
single-variable oscillators, we also observed that even (iii) networks
of diffusively coupled chaotic R\"{o}ssler systems \cite{Roessler}
behave similarly. On the same network topology, the dynamics of these
three kinds of distinct systems show closely related patterns of phase
disorder.

Commonly, transitions from synchrony to disordered dynamics have been
devoted to heterogeneities of, e.g., dynamical parameters or the degree
distribution, cf.\ \cite{Denker,Zumdieck,Motter}. \foreignlanguage{english}{}However,
the precise impact of topology onto the dynamics of directed
networks revealed here was so far not noticed. Even recent studies, considering
the exact dynamics of networks of given, specific topologies (see,
e.g.\ \cite{Earl,TimmePRL} and refs. therein) have not taken notice
of this impact. The main reason for this may be that all example networks
chosen to look at explicitely again were standard cases such as highly
connected random networks or lattices. 

Real-world oscillator networks, occurring across disciplines in physics,
biology and technology \cite{Strogatz,Acebron}, however, have a far
more complicated topology, and, as demonstrated in this Letter, may
thus strongly deviate in their dynamics. In a study \cite{BlankBunimovich}
related to ours, the notion of long range action has been introduced
showing that in certain directed networks of iterated maps the dynamics
of boundary units can control the dynamics of the entire network.
Our results suggest that local and global topological features,
such as the SCC super-network and the detailed topology of particular
SCCs, may act together to precisely control the dynamics of individual
units in complex directed networks. The concepts developed here
may thus also help to uncover information about the topologies of
such networks from their dynamics.

I thank M. Denker, M. Diesmann, T. Geisel, C. Kirst, A. Levina, R.-M.
Memmesheimer, F. Wolf, P. Ashwin, L. Bunimovich, J. Borresen, S. Gro\ss kinsky,
B. Kriener and S. Strogatz for stimulating discussions and comments
on the manuscript. I acknowledge partial support by the Federal Ministry
of Education and Research (BMBF), Germany, under grant number 01GQ0430.

\textbf{Appendix:} \emph{Level Structure of the SCC Super-Network:}
The level structure of the SCC super-network is constructed in three
steps. First, we determine the SCCs of the network using a standard
method \cite{Tarjan}, the computational complexity of 
which is $\mathcal{O}(N)$. Second, connections between them are straightforwardly derived from the underlying connections between units comprising these SCCs.
A connection from one SCC to another, $s\rightarrow s'$ is present
if there are $i\in s$ and $j\in s'$with a connection, $i\rightarrow j$
between them. Third, we find the longest undirected path from any
source SCC (without incoming connections) to any sink SCC (without
outgoing connections). The length of such a path is found by counting
a connection followed along its direction as {}``+1'' and against
its direction as {}``-1''. All units $i$ in a source SCC of the
longest path is given the level number $\ell(i)=1$. The levels of
all other SCCs are determined recursively according to the above counting
rule. The computational costs of finding the inter-SCC connections and the level structure strongly depend on the network under consideration.

\emph{Dynamics From Network Topology:} Given the level structure,
the linearized dynamics of every unit is determined for all units
in every given level, starting with level $\ell=1$ and proceeding
through subsequent levels recursively. Let $\boldsymbol{\phi}=\left(\boldsymbol{\phi}^{(1)},\ldots,\boldsymbol{\phi}^{(L)}\right)=\left(\phi_{i_{1}},\ldots,\phi_{i_{N}}\right)$
denote the asymptotic phases of all units $\phi_{i}$ in terms of
the collection of phases $\boldsymbol{\phi}^{(\ell)}$ of the units
at a given level $\ell\in\{1,\ldots,L\}$. For all units $i$ with
$\ell(i)=1$, their final states are $\phi_{i}^{(1)}=c_{s}$, where
$c_{s}$ depends on the initial state $\boldsymbol{\phi}_{s}(0)$
restricted to the SCC $s$. It equals the first component of the vector
$\mathbf{{c}}_{s}=V^{-1}\boldsymbol{\phi}_{s}(0)$, where $V=(\mathbf{{v}}_{i_{1}},\ldots,\mathbf{{v}}_{i_{R}})$
is a matrix of the $R$ eigenvectors $\mathbf{{v}}_{i_{r}}$ of $A$
localized on the SCC $s$ and $\mathbf{{v}}_{i_{1}}$ is the eigenvector
corresponding to the eigenvalue $\lambda_{i_{1}}=1$. This yields
the vector $\boldsymbol{\phi}^{(1)}$ of asymptotic phases in all
units in level $\ell=1$.

The phases $\boldsymbol{\phi}^{(\ell)}$ of units in the other levels
$\ell\geq2$ are determined iteratively given the phases $\boldsymbol{\phi}^{(\ell-1)}$
of units in level $\ell-1$. If some of the $\boldsymbol{\phi}^{(\ell)}$
are part of a source SCC in level $\ell$, these phases $\boldsymbol{\phi}_{\textrm{source}}^{(\ell)}$
are determined analogous to those in level $\ell=1$. The corresponding
sub-matrices $A_{\ell,\ell-1}$ and $A_{\ell,\ell}$ of the matrix
$A$ in (\ref{eq:linearmap}) needed to determine the remaining phases
$\boldsymbol{\phi}_{\textrm{no source}}^{(\ell)}$ describe the interactions
with units of the previous level $\ell-1$, and within the SCCs of
the current level $\ell$, respectively. Note that by definition of
the level structure, there are no interactions from level $\ell$
to level $\ell-1$. Thus the equation encoding this uni-directional
dependence, 
$\boldsymbol{\phi}_{\textrm{no source}}^{(\ell)}=A_{\ell,\ell-1}\boldsymbol{\phi}^{(\ell-1)}+A_{\ell,\ell}\boldsymbol{\phi}_{\textrm{no source}}^{(\ell)}$,
yields \begin{equation}
\boldsymbol{\phi}_{\textrm{no source}}^{(\ell)}=(1-A_{\ell,\ell})^{-1}A_{\ell,\ell-1}\boldsymbol{\phi}^{(\ell-1)}\label{eq:phi_no_source}\end{equation}
 such that, together with the $\boldsymbol{\phi}_{\textrm{source}}^{(\ell)}$
from above, all phases $\boldsymbol{\phi}^{(\ell)}$ of units in level
$\ell$ are determined. Iterating this for all levels $\ell\in\{2,\ldots,L\}$
we obtain the linear prediction of the complete disordered asymptotic
state $\boldsymbol{\phi}$. This analysis only
depends on the linearized effective couplings $A_{ij}$ that
determine the SCC super-network and  applies thus not restricted to the system (\ref{eq:KuramotoDE}).

\end{document}